\documentclass{llncs}
\pdfoutput=1
\usepackage{amssymb, amsmath, amsfonts}
\usepackage[T1]{fontenc}
\usepackage{lmodern} 

\begin{document}

\newtheorem{ttd}{Definition}
\newtheorem{ttt}{Theorem}
\newtheorem{ttl}{Lemma}
\newtheorem{tta}{Algorithm}
\newtheorem{ttc}{Corollary}
\newtheorem{ttp}{Property}

\mainmatter  

\title{Privacy by Design: On the Conformance Between Protocols and Architectures\protect\footnote{The final publication is available at link.springer.com (URL not yet available).}}

\titlerunning{Privacy by Design}
\author{Vinh-Thong Ta \and Thibaud Antignac}
\authorrunning{V.-T. Ta, T. Antignac}

\institute{\small INRIA, University of Lyon, France\\
\{vinh-thong.ta, thibaud.antignac\}@inria.fr}
%
%

\maketitle

\begin{abstract}
In systems design, we generally distinguish the architecture and 
the protocol levels. In the context of privacy by design, in the first case, 
we talk about privacy architectures, which define the privacy goals and 
the main features of the system at high level. In the latter case, we consider 
the underlying concrete protocols and privacy enhancing technologies  
that implement the architectures. In this paper, we address the question that 
whether a given protocol conforms to a privacy architecture and provide the 
answer based on formal methods. We propose a process algebra variant to define 
protocols and reason about privacy properties, as well as a mapping procedure 
from protocols to architectures that  are defined in a high-level architecture 
language. 
\end{abstract}

\section{Introduction} 
According to the definition provided in~\cite{bass:2012}, ``the architecture of a system is the set of [elements 
and their relations] needed to reason about the system''. In the context of privacy, the elements are typically 
the privacy enhancing technologies (PETs) themselves and the purpose of the architecture is to combine them to 
achieve the privacy requirements. Generally speaking, an architecture can be seen as the abstraction of a system 
since an architecture abstracts away the details provided by PETs (such as message ordering, timing, complex 
cryptographic algorithms, etc.). Architectures only capture the main functionalities that a system should 
provide, for instance, which computations and communications are to be performed by the components.

Works in privacy by design mainly focus on PETs rather than architectures.  
In the position paper \cite{Antignac:2014}, the authors addressed the problem of privacy by design at 
the architecture level and proposed the application of formal methods that facilitate a systematic architecture design. 
In particular, they provided the idea of \textit{the architecture language and logic}, a dedicated variant of epistemic logics~\cite{fagin:2004}, to deal with different aspects of privacy. Basically, an architecture is defined as a set of \textit{architecture relations}, which capture the computations and communications 
abilities of each component. For instance, a relation \textit{compute}$_{i}$($x$ $=$ $t$) specifies that 
a component $i$ can compute a value $t$ for $x$. 
Nevertheless, since \cite{Antignac:2014} is a position paper, the language envisioned in the paper is 
mainly based on an introductory description. An extended version of this language is detailed in 
\cite{TM-STM2014}. 

In this paper, we address the major question that whether the integration or combination of several 
different PETs conforms to a particular architecture. One challenge we have to face is that due 
to the diversity of technologies and protocols, their combination can raise a huge number of 
scenarios. Moreover, architectures are defined in an abstract way, while concrete 
implementations are more detailed, and it is challenging to define a proper abstraction 
from a lower to a higher level. The goal of this paper is to provide answers to this question.  

Specifically, our main contributions are two-fold: first, we propose a modified variant of 
the applied $\pi$-calculus \cite{fournet01mobile} for specifying the protocols related to PETs, 
and reasoning about the knowledge of components during the protocol run. Second, we propose a 
mapping procedure which defines the connection between the protocol specified in the calculus and 
the architecture defined in the architecture language. This mapping allows us to show whether a 
protocol (or a combination of protocols) conforms to a given architecture. To the best of our 
knowledge, this work is the first attempt that examines the connection between the two levels based 
on formal methods in the context of privacy protection. 

The paper is organized as follows: in Section \ref{sec:architectures}, we review the 
privacy architectures language (PAL) proposed in \cite{TM-STM2014}, which is a high-level language 
for specifying architectures and reasoning about privacy requirements in them. Sections \ref{sec:calc} 
and \ref{sec:extrarch} contain our contributions. The modified 
applied $\pi$-calculus is given in Section \ref{sec:calc}. Section \ref{sec:extrarch} 
discusses the connection  between each calculus process and relation in PAL, as well as the definitions 
and properties of the conformance between the two levels. In Section~\ref{related} we review the most 
relevant related works.  Finally, we conclude the paper and discuss about the future works in 
Section \ref{sec:conclusion}.

\section{Architecture Level}
\label{sec:architectures}

The language we review here is a simplified version of the one in \cite{TM-STM2014}.  
The functionality of a service is defined by $\Omega = \left\{ \tilde{X} = T \right\}$, where $T$ is a term 
and $\tilde{X} \in \textit{Var}$ represents a variable that can 
be either indexed ($X_K$) or unindexed ($X$). Each $\tilde{X}$ can be a single variable or an array of 
variables. $F \in Fun$ denotes a function, and 
$\odot F(X)$ defines the iterative application of $F$ to the variables in 
the array $X$ (e.g.: $\odot +(X)$ defines the summation of the variables in a given array). 

\small
\begin{tabbing}
    12345678901\= \kill
    \> $T$ ::= $\tilde{X}$ $|$ $F$($T_1$, \dots, $T_n$) $|$ $\odot$ $F$($X$);\ \ \ \ \ \ \ \ \ \ \ \ $\tilde{X}$ ::= $X$ $|$ $X_K$
\end{tabbing}
\normalsize

\small
\begin{tabbing}
    1234567890\=123\=1234\=12456789012345678901234567890123456789012345678901234567890123\= \kill
    \> $\mathcal{A}$ \> ::= \> \{$\mathcal{R}$\}\\
    \> $\mathcal{R}$ \> ::= \> $\textit{Has}^{arch}_i \left( \tilde{X} \right)$ $|$ $\textit{Receive}_{i,j} \left( \{\textit{Att}\}, \tilde{X} \right)$ $|$ $\textit{Compute}_i \left(\tilde{X} = T \right)$\\
    \>\>\> $|$ $\textit{Check}_i \left( \textit{T}_1 = \textit{T}_2 \right)$ $|$ $\textit{Verif}^{\textit{Attest}}_j \left( \textit{Att} \right)$ $|$ \textit{Trust}$_{i,j}$\\ 
    \> \textit{Att} \> ::= \> $\textit{Attest}_i \left( \{\tilde{X}=T\}\right)$
\end{tabbing}
\normalsize

An architecture $\mathcal{A}$ is 
defined by a set of components $C_i$, $i \in [1,\ldots,n]$,  
associated with the set of \textit{relations} $\{\mathcal{R}\}$. Each 
\textit{relation} $\mathcal{R}$ specifies a capability of the components. 
Subscripts $i$ and $j$ denote component IDs.  $\textit{Has}^{arch}_i(\tilde{X})$ expresses the 
fact that $\tilde{X}$ is a variable that component $C_i$ initially has (i.e., an input variable of $C_i$). 
$\textit{Receive}_{i,j}(\{\textit{Att}\}, \tilde{X})$ expresses the possibility for $C_i$ 
to receive the variable $\tilde{X}$ directly from $C_j$, and optionally an attestation $Att$
related to this variable. An attestation, defined by $\textit{Attest}_i(\{\tilde{X}=T\})$, 
captures a statement made by $C_i$ on the set of equations. Each $\tilde{X}$ $=$ $T$ in the 
set \{$\tilde{X}$$=$$T$\} expresses the integrity of $\tilde{X}$, stating that it equals to $T$. 
$\textit{Compute}_i(\tilde{X} = T)$ says that $C_i$ can compute a variable defined by an 
equation $\tilde{X}$ $=$ $T$. $\textit{Check}_i\left(T_1 = T_2\right)$ states 
that $C_i$ can check the satisfaction of property $T_1 = T_2$. The property 
$\tilde{X}$ $=$ $T$ in $\textit{Attest}_i$ is related to the same property in 
$\textit{Compute}_i$, namely when $C_i$ computes $\tilde{X}$ $=$ $T$ it can send an 
attestation on this. $\textit{Verif}^{\textit{Attest}}_j (\textit{Att})$ says 
that $C_j$ is able to successfully verify the origin of an attestation. Finally, 
${\textit{Trust}}_{i,j}$ is used to express the fact that component $C_i$ trusts $C_j$, and 
this trust relation does not change during operations. Trust relations are pre-defined, 
and an attestation sent by $C_i$ will be accepted by $C_j$ after a successful verification  
only if $C_j$ trusts $C_i$. 
 
The semantics of an architecture is defined as its sets of compatible traces. A trace is a 
sequence of possible high-level events occurring in the system. Events can be seen as 
\textit{instantiated relations} of the architecture. 

\small
\begin{tabbing}
    12345678\=123\=1234\=12456789012345678901234567890123456789012345678901234567890123\= \kill
    \> $\theta$ \> ::= \> $\textit{Seq}(\epsilon)$\\
    \> $\epsilon$ \> ::= \> $\textit{has}_i \left( \tilde{X}:V \right)$ $|$ $\textit{receive}_{i,j} \left( \{\textit{Att}\}, \tilde{X}:V \right)$ $|$ $\textit{compute}_i \left( \tilde{X} = T \right)$\\
    \>\>\> $|$ $\textit{check}_i \left(T_1 = T_2 \right)$ $|$ $\textit{verif}^{\textit{Attest}}_j \left( Att \right)$
\end{tabbing}
\normalsize

To distinguish events from relations, we let events start with lowercase. For instance, 
event $\textit{has}_i \left( \tilde{X}:V \right)$ captures the fact that $C_i$ has the value $V$ 
for $\tilde{X}$, and $\textit{compute}_i \left( \tilde{X} = T \right)$ expresses the fact that 
$C_i$ performes the computation $\tilde{X} = T$. The other events are interpreted based on their 
corresponding relations (see \cite{TM-STM2014} for details). 
An event trace $\theta$ is  \textit{compatible} with an architecture $\mathcal{A}$, 
if in this trace, only events which are instantiations of components of the architecture can 
appear in $\theta$ -- except for the \textit{compute} events. For the case of \textit{compute} events, besides the computation specified explicitly in the architecture, we also take into account the ``background'' computations (deduction) that can be performed by each component, based on the data it has. 
This deduction ability of each $C_i$ is captured by its deduction system $\triangleright_i$ \cite{TM-STM2014}. 
The semantics of events is based on the \textit{component states} and the \textit{global state} of 
the architecture, given as follows: 
\small
\begin{tabbing}
    1234567\=12345\=\kill
    \> \textit{State} \> $=$ \textit{State}$_V$ $\times$ \textit{State}$_P$;\\
    123456\=123456\=\kill
    \> \textit{State}$_V$ \> $=$ (\textit{Var} $\rightarrow$ \textit{Val}$_{\bot}$);\ \ \ \ \ \ \ \textit{State}$_{P}$ $=$ $\left\{\{\tilde{X} = T\} \cup \{T_1 = T_2\} \cup \{\textit{Trust}_{i,j}\}\right\}$ 
\end{tabbing}
\normalsize

The state of a component (\textit{State}) is composed of a variable state (\textit{State}$_{V}$) and a property state (\textit{State}$_{P}$). \textit{State}$_{V}$ assigns a value (which can be
undefined, $\bot$) to each variable. \textit{State}$_{P}$ defines the set of properties 
$\tilde{X} = T$ and $T_1 = T_2$ known by a component. 

In the sequel, $\sigma$ is used to denote the global state of the architecture 
$\mathcal{A}$ (state of the components $\langle C_1, \dots, C_n \rangle$) 
defined on $\textit{State}^n$. $\sigma_i$ ($\sigma_i$ $=$ ($\sigma^v_i$, $\sigma^{pk}_i$))  
denotes the state of the component $C_i$, where $\sigma^v_i$ and $\sigma^{pk}_i$ 
represent the variable state and property state of $C_i$, respectively.  
The initial state of $\mathcal{A}$, denoted by $\textit{Init}^\mathcal{A}$, contains 
only the trust properties specified by the architecture. 
The semantics of an event trace is defined by the function $\mathcal{S}_T$, which 
specifies the impact of a trace of events on the states of the 
components, through the impact of each event on the states (defined by the function $S_E$). 

\small
\begin{tabbing}
    123456\=12345678901\=123\= \kill
     \> $\mathcal{S}_T$ $:$ \textit{Trace} $\times$ \textit{State}$^n$ $\rightarrow$ \textit{State}$^n$;\ \ \ \ \ \ \ \ \ $\mathcal{S}_E$ $:$ \textit{Event} $\times$ \textit{State}$^n$ $\rightarrow$ \textit{State}$^n$;\\
    \> $\mathcal{S}_T$ ($\epsilon.\theta$, $\sigma$) $=$  $S_T$($\theta$, $S_E$($\epsilon$, $\sigma$));\ \ \ \ \ \ \ \ \ \    
    $\mathcal{S}_T$ ($\langle\rangle$, $\sigma$)  $=$ $\sigma$;\\ 
    \> $\mathcal{S}_E$ ($\textit{compute}_i \left( \tilde{X}= T \right), \sigma)$ $=$  $\sigma$[($\sigma^v_i$[$\textit{eval}(T,\sigma^v_i)$$/$$\tilde{X}$],
    $\sigma^{pk}_i$ $\cup$ \{$\tilde{X}$ $=$ $T$\}) $/$ $\sigma_i$ ].
\end{tabbing}
\normalsize 
 
Due to lack of space we only present $\mathcal{S}_E$ for the compute event here, 
the full list can be found in \cite{TM-STM2014}. The notation $\epsilon.\theta$ is used to 
denote a trace whose first element is $\epsilon$ and the rest of the trace is $\theta$, while 
$\langle\rangle$ denotes the empty trace. Each 
event modifies only the state $\sigma_i$ of the component $C_i$. A modification is expressed by 
$\sigma [(v, \textit{pk}) / \sigma_i]$ that replaces $\sigma^v_i$ and $\sigma^{pk}_i$ of $\sigma_i$ by $v$ 
and $\textit{pk}$, respectively. 
The effect of ${\textit{compute}}_i(\tilde{X}=T)$ is to set $\tilde{X}$ to the evaluation of 
$T$ based on the current variable state $\sigma^v_i$, which is denoted by 
$\textit{eval}(T,\sigma^v_i)$. Event \textit{compute}$_i$ also results in adding the knowledge 
about $\tilde{X} = T$ to the property state $\sigma^{pk}_i$. 
 
The semantics of an architecture $\mathcal{A}$ is defined as  
$\mathcal{S}(\mathcal{A}) = \{\sigma \in \textit{State}^n \, | \, \exists \theta \in T(\mathcal{A})\ \textrm{such that}\  S_T(\theta, \textit{Init}^\mathcal{A}) = \sigma \}$, where $T(\mathcal{A})$ is the set of compatible traces of $\mathcal{A}$.   
To reason about the privacy requirements of architectures, the \textit{architecture logic} is proposed in \cite{TM-STM2014}, which is based on the architecture language PAL.

\small
\begin{tabbing}
    1234567890\=12\=1234\=12456789012345678901234567890123456789012345678901234567890123\= \kill
    \> $\phi$ \> ::= \> $\textit{Has}^{all}_i\left( \tilde{X} \right)$ $|$ $\textit{Has}^{none}_i \left( \tilde{X} \right)$  $|$ $\textit{K}_i \left(\textit{T}_1 = \textit{T}_2 \right)$ $|$ $\phi_1 \wedge \phi_2$
\end{tabbing}
\normalsize

This logic involves modality $K_i$ that represents the epistemic knowledge~\cite{fagin:2004} 
of $C_i$ about \textit{T}$_1$ $=$ \textit{T}$_2$. In the rest of the paper, we 
refer to $\phi$ as an architecture property. 
The semantics $S(\phi)$ of a property $\phi$ is defined as follows: 
\small
\begin{tabbing}
    123\=\kill
    \>1. $\mathcal{A}$ $\in$ $S(\textit{Has}^{all}_i\left( \tilde{X} \right))$ $\Leftrightarrow$ $\exists$ $\sigma$ $\in$ $\mathcal{S}(\mathcal{A})$, $\sigma_i^v(\tilde{X})$ $\neq \bot$\\
    \>2. $\mathcal{A}$ $\in$ $S(\textit{Has}^{none}_i \left( \tilde{X} \right))$ $\Leftrightarrow$ $\forall$ $\sigma$ $\in$ $\mathcal{S}(\mathcal{A}), \sigma_i^v(\tilde{X})$ $=$ $\bot$\\ 
    \>3. $\mathcal{A}$ $\in$ $S(\textit{K}_i \left( \textit{Eq} \right))$  $\Leftrightarrow$ $\forall$ $\sigma^{\prime}$ $\in$ $\mathcal{S}_i(\mathcal{A})$, $\exists$ $\sigma$ $\in$ $\mathcal{S}_i(\mathcal{A})$, $\exists$ $\textit{Eq}^{\prime}$: ($\sigma$ $\geq_i$ $\sigma^{\prime}$) $\wedge$ ($\sigma_i^\textit{pk}$ $\triangleright_i$ $\textit{Eq}^{\prime})$\\ 
    \>\ \ \ \ \ \ \ \ \ \ \ \ \ \ \ \ \ \ \ \ \ \ \ \ \ \ \ \ \ \ \ \ \ \ \ \ \ \ \ \ \ \ \ \ \ \ \ \ \ \ \ \ \ \ \ \ \ \ \ \ \ \ \ \ \ \ \ \ \ \ \ \ \ \ \ $\wedge$ ($\textit{Eq}^{\prime}$ $\Rightarrow$ $\textit{Eq}$), 
\end{tabbing}
\normalsize

\noindent where \textit{Eq} (\textit{Eq}$'$) represents an equation $T_1$ $=$ $T_2$ ($T'_1$ $=$ $T'_2$). 
An architecture satisfies the $\textit{Has}^{all}_i(\tilde{X})$ property if and only if $C_i$ may 
obtain the value of all $X_k$ in \textit{Range}($X$) in at least one compatible execution trace. 
$\textit{Has}^{none}_i(\tilde{X})$ holds if and only if no execution trace can lead to a 
state in which $C_i$ gets any value of any $X_k$. We note that $\textit{Has}_i$ properties 
only inform on the fact that $C_i$ can get or derive some values for the variables but they 
do not bring any guarantee about the correctness of  these values. Integrity requirements 
can be expressed using the property $K_i(\textit{T}_1 = \textit{T}_2)$, which states that 
the component $C_i$ knows the truth of the integrity property 
$ \textit{T}_1 = \textit{T}_2$. 
In $\sigma \geq \sigma^{\prime}$, compared to $\sigma^{\prime}$, $\sigma$ represents the 
state at the end of a longer trace.
Finally, $\sigma_i^\textit{pk}$ $\triangleright_i$ $\textit{Eq}^{\prime}$ and $\textit{Eq}^{\prime}$ 
$\Rightarrow$ $\textit{Eq}$ capture that $\textit{Eq}^{\prime}$ can be 
deduced from $\sigma_i^\textit{pk}$  and $\textit{Eq}^{\prime}$, respectively.  

\textbf{Example Architecture}: Let us consider a very simple smart 
metering architecture which consists in the communication between two components: the meter ($M$) 
and the operator ($O$). The goal of this architecture is to ensure that the operator will get the consumption 
fee for a given period and to be convinced that the fee is correct. The privacy requirement 
says that $O$ must not obtain the consumption data. One possible design solution is that the meter passes directly the consumption data to the operator who will compute the fee:  

\small
\begin{center}
$\mathcal{A}_1$ $=$ \{for $i$ $\in$ [$1$, \dots, $r$]: \textit{Has}$^{arch}_M$($X_{c}$); \textit{Compute}$_M$($X_{m_i}$ $=$ $X_{c_i}$); 
\textit{Receive}$_{O, M}$(\textit{Attest}$_{M}$($X_{m_i}$ $=$ $X_{c_i}$), $X_{m_i}$); 
\textit{Verif}$^{Attest}_{O}$(\textit{Attest}$_{M}$($X_{m_i}$ $=$ $X_{c_i}$)); 
\textit{Compute}$_O$($X_{\textit{tf}_i}$ $=$ $F(X_{m_i}$); \textit{Compute}$_O$($X_{\textit{fee}}$ $=$ 
$\odot +(X_{\textit{tf}})$), \textit{Trust}$_{O, M}$\}. 
\end{center}
\normalsize

In the architecture $\mathcal{A}_1$, the meter initially has the (input) variable $X_{c}$ that 
represents the array of $r$ consumption data $X_{c_i}$, $i$ $\in$ [$1$, \dots, $r$]. 
The meter is capable to compute each metered data ($X_{m_i}$) based on each consumption data ($X_{c_i}$). Intuitively, 
in $X_{m_i}$ $=$ $X_{c_i}$, $X_{m_i}$  will get the value 
of $X_{c_i}$ . Then, the operator will receive the metered data ($X_{m_i}$), 
along with the attestation made by $M$ on the integrity property $X_{m_i}$ $=$ 
$X_{c_i}$. After verifying the received attestation with success, due to \textit{Trust}$_{O, M}$ 
the operator knows that  $X_{m_i}$ $=$ $X_{c_i}$. Then for each 
$X_{m_i}$, $O$ computes the tariff based on the function $F$. Finally, $O$ computes 
the summation of the $r$ tariffs (i.e., array $X_{\textit{tf}}$) to get the fee for the period. 
The requirements of the architecture are modeled with the properties of the architecture logic. Namely, 
$\textit{Has}^{all}_O\left( X_{fee} \right)$ specifies that $O$ has (all) the fee, while 
$\textit{Has}^{none}_O\left( X_{c} \right)$ says that $O$ must not have any consumption data. 
$\mathcal{A}_1$ fulfills the first requirement, but it does not satisfy the privacy requirement because  
based on \textit{Receive}$_{O, M}$(\textit{Attest}$_{M}$($X_{m_i}$ $=$ $X_{c_i}$ ), 
$X_{m_i}$) $O$ can obtain $X_{c_i}$ from $X_{m_i}$. 

\section{Protocol Level} 
\label{sec:calc}

To reason about the concrete implementations of an architecture, we propose a modified variant of the applied $\pi$-calculus. 
We decided to modify the basic applied $\pi$-calculus  \cite{fournet01mobile} because thanks to its expressive syntax 
and semantics, it is broadly used for security verification of systems and protocols (e.g., 
\cite{Fournet:2002}, \cite{Delaune:2008}, \cite{DongJP:2010}, \cite{Li:2009}, \cite{Delaune:2009}, \cite{BackesZero:2008}, 
\cite{Kremer:2005}).  
Our main goal is to modify some syntax and semantics elements of the applied $\pi$-calculus, 
making it more convenient to find the connection between the calculus semantics and the interpretation of 
architecture relations. One of such modifications is the notion of component, which is characterized 
by three elements: (i) the internal behavior of the component; (ii) the unique ID assigned to the component; and (iii) 
the set of IDs of the components who are trusted by this component. Another reason why we cannot use 
the basic applied $\pi$-calculus is that it focuses on reasoning about the information a 
Dolev-Yao attacker (who can eavesdrop on all communications) obtains. However, in our case we reason about 
the information that components can have, which are only 
aware of the communications they can take part in.

\subsection{Syntax of the Modified Applied $\pi$-Calculus}
\label{sec:calcsyntax}
We assume an infinite set of \textit{names} $\mathcal{N}$ and  
\textit{variables} $\mathcal{V}$, and a finite set of  \textit{component identifiers} 
$\mathcal{L}$, where $\mathcal{V}$ $\cap$ $\mathcal{N}\cap \mathcal{L} =\emptyset$. 
Terms are defined as follows:
\begin{tabbing}
    123456789012\=12\=1234\=\kill
    \> $t$ \> ::= \> $c$\ \ $|$\ \  $l_i$\ \   $|$\ \  $n, m, k$\ \ $|$\ \  $x,y,z$\ \  $|$\ \  $f(t_1,\dots, t_p)$. 
\end{tabbing}
  
\noindent  The meaning of each term is given as follows:   
$c$ models a communication channel.   
$l_i$ represents a component ID ($l_i$ $\neq$ $l_j$ if $i$ $\neq$ $j$) that uniquely identifies a component. 
$n$, $m$ and $k$ denote names, which model some kind of data (e.g., a random nonce, a secret key, etc.).  
Terms $x$, $y$, $z$ denote variables that represent any term, namely, 
any term can be bounded to variables. 
$f(t_1,\dots, t_p)$ is a function, which models 
cryptographic primitives, e.g., digital signature 
can be modeled by $\textit{\textrm{sign}}(x_m, x_{sk})$, where $x_m$ 
and $x_{sk}$ specify the message and the private key, respectively. Moreover, $f$ 
can also be used to specify verification functions (e.g., the signature 
check is modeled by function 
\textit{checksign}(\textit{sign}($x_m$, $x_{sk}$), $x_{pk}$), where 
$x_{pk}$ represents the public key corresponding to the private key $x_{sk}$).  
  
We rely on the same type system for terms as in the applied $\pi$-calculus \cite{fournet01mobile}. 
Due to lack of space, we omit the unimportant details of this type system, and 
leave it implicit in the rest of the paper. We assume that terms 
are well-typed and that substitutions preserve types (see \cite{ryan:2011} 
for details).

The internal operation of components is modeled by \textit{processes}. Processes 
are specified with the following syntax:

\small
\begin{tabbing}
    \=\kill
    \> $P$, $Q$, $R$ ::=\ \ $\overline{c}\langle t \rangle.P$\ \ \ $\|$\ \ \ $\overline{c}\langle t_{m}, t_{sig}) \rangle.P$\ \ \ $\|$\ \ \ $c(x)$.$Q$\ \ \ $\|$\ \ \ $c(x_m, x_{sig})$.$Q$\ \ \ $\|$\ \ \ $P | Q$\\
    \>\ \ \ \ \ \ \ \ \ \ \ \ \ \ \ \ \ \ $\|$\ \ \ $\nu n.P$\ \ \ $\|$\ \ \  \textit{let} $x$ $=$ $t$ \textit{in} $P$\ \ \ $\|$\ \ \textit{if} $(t_1 = t_2)$ \textit{then} $P$\ \ $\|$\ \ \ \textbf{0}.   
\end{tabbing}
\normalsize

Note that for simplicity we left out the infinite replication of processes, $!P$. As a result 
a protocol/system run consists of a finite number of traces. 

Process $\overline{c}\langle t \rangle.P$ sends the term $t$ (where 
$t$ $\neq$ ($t_{m}$, $t_{sig}$)) on channel 
$c$, and continues with the execution of $P$.  
Process $\overline{c}\langle t_{m}, t_{sig} 
\rangle.P$ models the attestation sending, where $t_{m}$ 
and the signature $t_{sig}$ are sent on $c$. 

Process $c(x)$.$Q$ waits for a term on channel $c$ and then binds the 
received term to $x$ in $Q$.   
Process $c(x_m, x_{sig})$.$Q$ waits for a term $x_m$ and its signature 
$x_{sig}$ on channel $c$, which models the attestation reception. 

$P | Q$ behaves as processes $P$ and $Q$ running (independently) in
parallel.     
A restriction $\nu n.P$ is a process that creates a new, bound name
$n$, and then behaves as $P$. The name $n$ is called bound because it is available 
only to $P$.    
Process  \textit{let} $x$ $=$ $t$ \textit{in} $P$ proceeds to $P$ and binds
every (free occurrence of) $x$ in $P$ to $t$. 
 
Process \textit{if} $(t_1 = t_2)$ \textit{then} $P$ says that if 
$t_1$ $=$ $t_2$ (with respect to the equational theory $E$, discussed later) 
then process $P$ is executed, else it stops.  
Its special case is \textit{if} $x_m$ $=$ \textit{checksign}($x_{sig}$, $x^{pk}_{l_i}$) \textit{then} $P$, which  
captures the verification of an attestation (i.e., signature $x_{sig}$ with key $x^{pk}_{l_i}$). For message authentication and integrity protection purposes digital 
signature and message authentication code (MAC) are used. In this paper we 
only consider signature. 

Finally, the \textit{nil} process \textbf{0} does nothing and specifies process 
termination. 

\textbf{Components:} To make the connection between calculus processes and architecture 
relations more straightforward, we introduce the notion of components. 
$\left\lfloor P \right\rfloor^{\rho}_l$ defines a component with the unique identifier 
$l$, who trusts the components whose IDs are in the set $\rho$, and whose behavior is 
defined by process $P$. The trust relation can be either one-way or symmetric, for instance, 
$\left\lfloor P \right\rfloor^{\{l_2\}}_{l_1}$ and $\left\lfloor Q \right\rfloor^{\{l_1\}}_{l_2}$ 
represent components $l_1$ and $l_2$ who trust each other. The rationale behind this way of 
component specification is that the component IDs and the trust relation between them are 
pre-defined, and do not change during the protocol run (this is what we assumed at 
the architecture level). In addition, we assume that a trusted component will not become untrusted. 

\textbf{Systems:}
A \textit{system}, denoted by $S$, can be an empty system with no component: $\textbf{0}_S$; 
a singleton system with one component: $\left\lfloor P \right\rfloor^{\emptyset}_l$; the parallel 
composition of components: $\left\lfloor P \right\rfloor^{\rho_1}_{l_1}$ $|$ 
$\left\lfloor Q \right\rfloor^{\rho_2}_{l_2}$, where $\rho_1$ and $\rho_2$ 
may include $l_2$ and $l_1$, respectively; or a system with name restriction. To capture 
more complex systems, we also allow systems to be the parallel composition of sub-systems, 
$S_1$ $|$ $S_2$.
\begin{center}    
$S$ ::= $\textbf{\textrm{0}}_S$ $|$\ $\left\lfloor P \right\rfloor^{\rho}_{l}$ $|$\ $\nu n.S$ $|$\  $(S_1$ $|$ $S_2)$. \end{center}

\noindent 
The name restriction $\nu n.S$ represents the creation of new name $n$, such as secret keys, or a random 
nonce which are only available to the components in $S$. 

\subsection{Semantics of the Modified Applied $\pi$-Calculus}
\label{sec:calcsemantics}

In order to check the conformance between protocols and architectures, it suffices to consider  
the internal reduction rules of the calculus, which model the behavior of the protocol 
(without contact with its environment). Reduction rules capture the internal operations 
(e.g., \textit{let} or \textit{if} processes) and communications performed by components. We define 
and distinguish the following reduction rules:  
\small 
\begin{tabbing}
    \=123456789012345\=12345\=123\= \kill
    \> \textbf{(Reduction rules)}\\ 
    \=123456789\=1234567891234567890123\=12345678901\= \kill
    \> (Rcv) \> \small $\left\lfloor \overline{c}\langle t \rangle.P \right\rfloor^{\rho_i}_{l_i}$ $|$ $\left\lfloor c(x).Q \right\rfloor^{\rho_j}_{l_j}$ \> $\stackrel{rcv(l_j,l_i, x:t)}{\longrightarrow}$ \> $\left\lfloor P \right\rfloor^{\rho_i}_{l_i}$ $|$ $\left\lfloor Q \{t/x\}\right\rfloor^{\rho_j}_{l_j}$, $t$ $\neq$ ($t_m$, $t_{sig}$);\\\\
    \=123456789\= \kill
    \> (Rcv$_{att}$) \> $\left\lfloor \overline{c}\langle t_m, t_{sig} \rangle.P \right\rfloor^{\rho_i}_{l_i}$ $|$ $\left\lfloor c(x_{m}, x_{sig}).Q \right\rfloor^{\rho_j}_{l_j}$\ \ \  $\stackrel{rcv_{att}(l_j,l_i, x_m:t_m)}{\longrightarrow}$\\ 
    \>\ \ \ \ \ \ \ \ \ \ \ \ \ \ \ \ \ \ \ \ \ \ \ \ \ \ \ \ \ \ \ \ \ \ \ \ \ \ \ \ \ \ \ \ \ \ \ \ \ \ \ \ \ \ \ \ \ $\left\lfloor P \right\rfloor^{\rho_i}_{l_i}$ $|$ $\left\lfloor Q \{t_m/x_{m}, t_{sig}/x_{sig}\}\right\rfloor^{\rho_j}_{l_j}$;\\\\
    \=123456789\=12345678901234567890\=12345678901\= \kill
    \> (Verif$_{att}$) \> $\lfloor \textit{if}\ \ x_m\ =\  \textit{checksign}(x_{sig}, t^{pk}_{l_i})\ \textit{then}\ Q' \rfloor^{\rho_j}_{l_j}$\ \ \ $\stackrel{ver_{att}(l_j,\ x_m:t_m)}{\longrightarrow}$ $\lfloor Q' \rfloor^{\rho_j}_{l_j}$,\\ 
    \>\ \ \ \ \ \ \ \ \ \ \ \ \ where \{$t_m/x_{m}$, $t_{sig}/x_{sig}$\}. Note: $l_j$ accepts the attestation if $l_i$ $\in$ $\rho_j$.\\\\
    \=123456789\=123456789012345678901\=123456789012\= \kill 
    \>(Check) \> $\left\lfloor \textit{\textrm{if}}\ (t_1 = t_2)\ \textit{\textrm{then}}\ P \right\rfloor^{\rho_j}_{l_j}$ \> $\stackrel{\textit{check}(l_j, t_1:t_2)}{\longrightarrow}$ \> $\left\lfloor P \right\rfloor^{\rho_j}_{l_j}$ ($t_1 = t_2$ $\in$ $E$, $t_2$ $\neq$ \textit{checksign});\\\\  
    \=123456789\=1234567891234678901\=123456\= \kill
    \>(Comp) \> $\left\lfloor let\ x = t\ \ in\ P \right\rfloor^{\rho_j}_{l_j}$\> $\stackrel{\omega_{comp}}{\longrightarrow}$ \> $\left\lfloor P\{t/x\} \right\rfloor^{\rho_j}_{l_j}$, ($t$ $=$ $x'$ or $f$, such that\\ 
    \>\ \ \ \ \ \ \ \ \ \ \ \ \ \ $\omega_{comp}$ $=$  \textit{comp}($l_j$,\   $x:t$) when $f$ $\notin$ \{\textit{sign}, \textit{checksign}\}, else $\omega_{comp}$ $=$ $\tau$);\\\\  
    \=123456789\=1234567891234678901234\=1234567890\= \kill
    \> (Has) \> ($\nu k$.) $\left\lfloor let\ x = k\ in\ P \right\rfloor^{\rho_j}_{l_j}$\> $\stackrel{has(l_j, x:k)}{\longrightarrow}$ \> ($\nu k$.) $\left\lfloor P\{k/x\} \right\rfloor^{\rho_j}_{l_j}$,\\\\
    \=123456789\=1234567890123456701234\=123456\= \kill
    \>(Error) \> $\left\lfloor \textit{\textrm{if}}\ (t_1 = t_2)\ \textit{\textrm{then}}\ P \right\rfloor^{\rho_j}_{l_j}$ \> $\stackrel{\textit{error}}{\longrightarrow}$ \> $\left\lfloor \textbf{0} \right\rfloor^{\rho_j}_{l_j}$\ \ \ \ (\textit{if} $t_1 = t_2$ $\notin$ $E$);\\\\
   \=123456789\=123456789012345670123\=123456\= \kill
    \>(Par-C) \> $\left\lfloor P \right\rfloor^{\rho_j}_{l_j}$\ \ $\stackrel{\omega_c}{\longrightarrow}$\ \ $\left\lfloor P' \right\rfloor^{\rho_j}_{l_j}$ then  $\left\lfloor Q\ |\ P \right\rfloor^{\rho_j}_{l_j}$\ \ $\stackrel{\omega_c}{\longrightarrow}$\ \ $\left\lfloor Q\ |\ P' \right\rfloor^{\rho_j}_{l_j}$;\\\\ 
    \>(Res-C) \> $\left\lfloor P \right\rfloor^{\rho_j}_{l_j}$\ \ $\stackrel{\omega_c}{\longrightarrow}$\ \ $\left\lfloor P' \right\rfloor^{\rho_j}_{l_j}$ then  $\left\lfloor \nu n.P \right\rfloor^{\rho_j}_{l_j}$\ \ $\stackrel{\omega_c}{\longrightarrow}$\ \  $\left\lfloor \nu n.P' \right\rfloor^{\rho_j}_{l_j}$, where\\
    \>\ \ \ \ \ \ \ \ \ \ \ \  \ \ $\omega_c$ $\in$ \{\textit{comp}($l_j$, $x:t$), \textit{has}($l_j$, $x:k$), \textit{check}($l_j$, $t_1:t_2$), \textit{error}, \textit{ver}$_{att}$($l_j$, $x_m$$:$$t_m$)\};\\\\ 
    \>(Par-S) \> $S_1$\ \ $\stackrel{\omega_s}{\longrightarrow}$\ \ $S'_1$ then  $S_2$ $|$ $S_1$\ \ $\stackrel{\omega_s}{\longrightarrow}$\ \ $S_2$ $|$ $S'_1$;\\\\ 
    \>(Res-S) \> $S$\ \ $\stackrel{\omega_s}{\longrightarrow}$\ \ $S'$ then  $\nu n.S$\ \ $\stackrel{\omega_s}{\longrightarrow}$\ \ $\nu n.S'$, where\\
    \>\ \ \ \ \ \ \ \ \ \ \ \ \ \ $\omega_s$ can be $\omega_c$, and \textit{rcv}($l_j$, $l_i$, $x:t$), \textit{rcv}$_{att}$($l_j$, $l_i$, $x_m:t_m$).
\end{tabbing}
\normalsize

\noindent Before defining the system states, we label each reduction relation 
(arrow) based on the name of the rule and the terms used in them. 
We adopt the notion of equational theory $E$ from \cite{fournet01mobile}, \cite{ryan:2011}, 
which contains rules of form $t_1$ $=$ $t_2$ $\in$ $E$, that define when two terms are equal. For instance,  the 
equational theory $E$ may include rules for signature verification, decryption, MAC verification, etc. The 
meaning of each reduction rule is as follows: 

\begin{itemize}
\item 
Rule (Rcv) captures the communication between components $l_i$ and $l_j$. 
Namely, $l_i$ sends value $t$ for $x$ on channel $c$, which is received by $l_j$. As a result, 
we get $Q$$\left\{t / x\right\}$ that binds $t$ to every free occurrence of $x$ in $Q$. It is 
assumed that $t$ $\neq$ ($t_m$, $t_{sig}$), which is treated as a special case. 
\item 
Rule (Rcv$_{att}$) deals with exchanging the message $t_{m}$ and the signature 
$t_{sig}$ on channel $c$, which models the reception of the 
attestation \textit{Attest}$_{l_i}$(\{$x_m$ $=$ $t_m$\}). As a result, we get $Q$ in which 
$t_{m}$ and the signature are bound to $x_m$ and $x_{sig}$, respectively. The reason that 
we distiguish (Rcv$_{att}$) from (Rcv) is because we want to make a clear distinction 
between the cases of receiving a message with and without an attestation. 

\item 
Rule (Verif$_{\textit{att}}$) captures the case when after binding $t_m$ and $t_{sig}$ to 
$x_m$ and the signature $x_{sig}$, respectively, component $l_j$ successfully verified 
the signature using the corresponding public key of $l_i$, $t^{pk}_{l_i}$. 
We implicitly assume that $t_{m}$ and $t_{sig}$ contain enough information for the 
receiver to identify the ``type'' of the received message (e.g., the consumption fee in smart metering systems). 
Rules (Rcv$_{\textit{att}}$) and (Verif$_{\textit{att}}$) together specify the scenario when 
$l_i$ sends to $l_j$ the value $t_{m}$ for $x_m$, with the signature that proves the integrity 
and the authenticity of  this message. Then, in case $l_j$ trusts $l_i$, it knows 
the truth about the integrity property $x_m$ $=$ $t_{m}$.

\item 
Rule (Check) considers the case when two terms are equal in the check 
(with respect to the equational theory $E$), which leads to $P$ as result. We assume that 
$t_2$ is not the \textit{checksign} function, which is used for the attest verification.   

\item 
Rule (Comp) models the computation $x$ $=$ $t$ performed by $l_j$.
As a result, every free occurrence of $x$ in $P$ is given the value $t$. 
In this rule we assume that $t$ can be either a variable or a function (except for 
\textit{sign} and \textit{checksign}, because they are considered as parts of the attestation), 
but not a name. 

\item 
Rule (Has) deals with the case when $t$ is a name. 
The name $k$, either bound (with $\nu k$) or free (without $\nu k$), represents the value of $x$. Here 
$x$ is used to model the variable that $l_j$ initially has to capture the input data 
coming from the environment (e.g., the consumption data in the smart metering).    

\item 
Rule (Error) specifies the case when two terms are not equal with respect to $E$. As a result, 
the process will continue with the \textit{nil} process. 

\item 
Rules (Par-C) and (Res-C) say that the \textit{if} and \textit{let} reductions are 
closed under parallel composition and restriction within a component, respectively.   
Rules (Par-S) and (Res-S) capture that all the reductions are 
closed under parallel composition and restriction on systems, respectively.

\end{itemize}

Instead of referring to the trace of reductions $\stackrel{\omega^1_s}{\longrightarrow}$ \dots 
$\stackrel{\omega^m_s}{\longrightarrow}$ we will refer to the trace of the corresponding labels 
$\omega^1_s$, \dots, $\omega^m_s$ for the sake of clarity.         
 
\textbf{States of components and systems}: Let us consider a system $S$ with $n$ components. 
Let \textit{Label}$_S$ be the set of all labels ($\omega_s$ $\in$  \textit{Label}$_S$) of the reduction 
relations defined above, and let \textit{LTrace}$_S$ be the set of all possible label traces of $S$. 
We define the functions $\mathcal{V}_{\textit{ST}}$, $\mathcal{V}_{\textit{T}}$ and $\mathcal{V}_{L}$ 
that update the states of the components and the entire system. $\mathcal{V}_{\textit{T}}$ and 
$\mathcal{V}_{L}$ are similar to $\mathcal{S}_{T}$ and $\mathcal{S}_E$ in PAL, but they are based 
on label traces and labels instead of traces of events and events. $\mathcal{V}_{\textit{ST}}$ takes 
as input the set of all the possible label traces of $S$ and handle each trace with $\mathcal{V}_{\textit{T}}$.
Let \textit{State}$^n_S$ denote the 
set of global state of $S$ and $\emptyset_{tr}$ an empty set of traces. Finally, in 
$\omega_s$.\textit{tr} the label $\omega_s$ is the prefix of the trace \textit{tr}.

\small
\begin{tabbing}
    \=12345678901\=123\= \kill
    \> $\mathcal{V}_{ST}$ $:$ \{\textit{LTrace}$_S$\} $\times$ \textit{State}$^n_S$ $\rightarrow$ \textit{State}$^n_S$;\\
    \> $\mathcal{V}_{T}$ $:$ \textit{LTrace}$_S$ $\times$ \textit{State}$^n_S$ $\rightarrow$ \textit{State}$^n_S$; \ \ \ \ \ \ \    \ \ \ \ \ \ \  \ \ \ \ \ \ \  \ \ \ \ \ \ $\mathcal{V}_L$ $:$ \textit{Label}$_S$ $\times$ \textit{State}$^n_S$ $\rightarrow$ \textit{State}$^n_S$;\\ 
    \> $\mathcal{V}_{ST}$(\textit{LTrace}$_S$, $\lambda$)  $=$ $\mathcal{V}_{ST}$ (\textit{LTrace}$_S$$\backslash$\{\textit{tr}\}, 
    $\mathcal{V}_{T}$(\textit{tr}, $\lambda$)); \ \ \ \  
    $\mathcal{V}_{ST}$($\emptyset_{tr}$, $\lambda$)  $=$ $\lambda$;\\  
    \> $\mathcal{V}_{T}$($\omega_s$.\textit{tr}, $\lambda$) $=$  $\mathcal{V}_{T}$(\textit{tr}, $\mathcal{V}_L$($\omega_s$, $\lambda$)); \ \ \ \ \ \ \ \ \ \ \ \ \ \ \ \ \ \ \ \ \ \ \ \ \ \ \ $\mathcal{V}_{T}$ ($\left\langle \right\rangle$, $\lambda$) $=$ $\lambda$;\\
    \> $\mathcal{V}_L$($has(l_i, x:k)$, $\lambda$) $=$ $\lambda$[($\lambda^v_{l_i}$\{$k$/$x$\}, $\lambda^{pk}_{l_i}$) $/$ $\lambda_i$];\\
    \> $\mathcal{V}_L$($rcv(l_i,l_j, x:t), \lambda)$ $=$ $\lambda$[($\lambda^v_{l_i}$\{$t$/$x$\}, $\lambda^{pk}_{l_i}$) $/$ $\lambda_i$]\\
    \> $\mathcal{V}_L$($rcv_{att}(l_i,l_j, x_m:t_m)$, $\lambda$) $=$ $\lambda$[($\lambda^v_{l_i}$\{$t_m$/$x_m$\}, $\lambda^{pk}_i$) $/$ $\lambda_{l_i}$]\\
    \> $\mathcal{V}_L$($\textit{comp}(l_i, x:t)$, $\lambda)$ $=$  $\lambda$[($\lambda^v_{l_i}$\{$\lambda^v_{l_i}$$t$$/$$x$\}, $\lambda^{pk}_{l_i}$ $\cup$ \{$x$ $=$ $t$\}) $/$ $\lambda_{l_i}$]\\ 
    \> $\mathcal{V}_L$($\textit{check}(l_i, t_1:t_2)$, $\lambda$) $=$ $\lambda$[($\lambda^v_{l_i}$, $\lambda^{pk}_{l_i}$ $\cup$ \{$t_1$ $=$ $t_2$\}) $/$ $\lambda_{l_i}$] 
     if $\lambda^v_{l_i}$$t_1$ $=$ $\lambda^v_{l_i}$$t_2$ $\in$ $E$\\
    \> $\mathcal{V}_L$($ver_{att}(l_i,\ x_m:t_m)$, $\lambda$) $=$ $\lambda$[($\lambda^v_{l_i}$, $\lambda^{pk}_{l_i}$ $\cup$ \{\{$x_m$ $=$ $t_m$\} if  \textit{Trust}$_{l_i,l_j}$ $\in$  $\lambda^{pk}_{l_i}$\}) $/$ $\lambda_i$].
\end{tabbing}
\normalsize

We let $\lambda_{l_i}$ and $\lambda$ denote the state of component $l_i$ and   
the global state that consists in the state of all components in the system, respectively.  
Each $\lambda_{l_i}$ is defined by the pair ($\lambda^{v}_i$, $\lambda^{pk}_{l_i}$), which is the 
variable state and the property state for component $l_i$, respectively. In our calculus 
the variable state $\lambda^v_{l_i}$ is defined by the set of substitutions \{$t_1$$/$$x_1$, 
\dots, $t_m$$/$$x_m$\}, which captures the terms available to $l_i$, as well as the values of each 
variable from the perspective of $l_i$. $\lambda^v_{l_i}$\{$t$/$x$\} is a shorthand for 
($\lambda^v_{l_i}$ $\cup$ \{$t$/$x$\}) $\backslash$ \{$t'$/$x$\}, if \{$t'$/$x$\} $\in$ 
$\lambda^v_{l_i}$ for some $t'$. $\lambda^{pk}_i$ is the set of integrity properties (e.g., $t_1$ $=$ $t_2$) 
that captures the knowledge gained by $l_j$ about these properties.  
$\lambda^v_i$$t$ represents the evaluation of $t$ based on $\lambda^v_i$, and 
$\lambda^v_{l_i}$$t_1$ $=$ $\lambda^v_{l_i}$$t_2$ $\in$ $E$ says that the evaluation of 
$t_1$ and $t_2$ in $\lambda^v_{l_i}$ are equal up to the equational theory $E$.  
We also consider the state update that results after a failed check 
(namely, $\lambda$[$\lambda_{Err}$/$\lambda_i$], where $\lambda_{Err}$ denotes the error state), 
though we omit the formal details here to save space.  

\section{From Protocols to Architectures}
\label{sec:extrarch}   
In the sequel, we discuss how the corresponding architecture can be extracted based on a given protocol or system. 
Namely, given a protocol specified in our process calculus we define an extraction procedure that extracts 
the corresponding architecture relations. The extraction procedure is based on the application of a set of 
extraction rules that we define below. Each extraction rule specifies the connection between the traces of labels    
of a system and the corresponding architecture relation. We assume a (initial) system $S$ which consists 
in the parallel composition of $r$ components (for a finite $r$), namely, $S$ $\stackrel{def}{=}$ $\left\lfloor P_1 \right\rfloor^{\rho_1}_{l_1}\ |\ \dots \ |\ \left\lfloor P_r \right\rfloor^{\rho_r}_{l_r}$. 
The corresponding architecture relations will be extracted based on the possible traces (i.e., the trace semantics) 
of $S$. We emphasize that during the extraction of architectural properties we only consider the reduction traces to 
capture the communication between components, without considering the activity of the environment (i.e., the Dolev-Yao 
attacker). Formally, we do not take into account the labelled transitions known in the applied $\pi$-calculus \cite{fournet01mobile}. The 
reason is that the architecture relations focus only on the abilities of the components and the communication 
between them.

An architecture does not not contain the \textit{Compute} relations for background computations. 
The situation is similar at the protocol level, where the protocol description specifies 
the basic computations and communications of the components, without involving the background computations. 
Hence, when extracting the architecture relations, it is sufficient to consider only the protocol description and its
corresponding reduction traces. The background computations will be taken into account when we discuss the 
mapping to the \textit{Has}$_j$ architecture logic property for reasoning about the data that 
can be deduced by a component. 

Given a system $S$ and the set \textit{LTrace}$_S$ of (all) its possible label traces, we define the 
extraction function $\mathcal{X}_{T}$ that extracts the corresponding architecture based on 
\textit{LTrace}$_S$. $\mathcal{X}_{\textit{ST}}$ is interpreted similarly as $\mathcal{V}_{\textit{ST}}$. 
\textit{Rel}$_{S}$ denotes the set of architectural relations of $S$. Function $\mathcal{X}_L$ extracts 
a relation based on a label $\omega_s$ and put it into $\alpha_S$. We use $\alpha_S$ to denote the set 
of the extracted relations so far. We have the following 
extraction rules:  
\small
\begin{tabbing}
    \=12345678901\=123\= \kill
    \> $\mathcal{X}_{ST}$ $:$ \{\textit{LTrace}$_S$\} $\times$ \textit{Rel}$_{S}$ $\rightarrow$ \textit{Rel}$_{S}$;\\
    \> $\mathcal{X}_{T}$ $:$ \textit{LTrace}$_S$ $\times$ \textit{Rel}$_{S}$ $\rightarrow$ \textit{Rel}$_{S}$; \ \ \ \ \ \ \ \ \ \ \ \ \ \ \ \ \ \ \ \ \ \ \ \ \ \ \ \ \ \ \ \ \ \ \ \ $\mathcal{X}_L$ $:$ \textit{Label}$_S$ $\times$ \textit{Rel}$_{S}$ $\rightarrow$ \textit{Rel}$_{S}$;\\ 
    \> $\mathcal{X}_{ST}$(\textit{LTrace}$_S$, $\alpha_S$) $=$  $\mathcal{X}_{ST}$(\textit{LTrace}$_S$$\backslash$\{\textit{tr}\}, $\mathcal{X}_{T}$(\textit{tr}, $\alpha_S$));\ \ \ \ \ \  
    $\mathcal{X}_{ST}$($\emptyset_{tr}$, $\alpha_S$)  $=$ $\alpha_S$;\\  
    \> $\mathcal{X}_{T}$($\omega_s$.\textit{tr}, $\alpha_S$) $=$  $\mathcal{X}_{T}$(\textit{tr}, $\mathcal{X}_L$($\omega_s$, $\alpha_S$));\ \ \ \ \ \ \ \ \ \ \ \ \ \ \ \ \ \ \ \ \ \ \ \ \ \ \ \ $\mathcal{X}_{T}$ ($\left\langle \right\rangle$, $\alpha_S$) $=$ $\alpha_S$;\\
    \>$R^{\textit{has}}$:\ \ \ $\mathcal{X}_L$($has(l_j, x:k)$, $\alpha_S$) $=$ $\alpha_S$ $\cup$ \{\textit{Has}$^{arch}_{l_j}$($x$)\};\\
    \>$R^{\textit{recv}}$:\ \ \ $\mathcal{X}_L$($rcv(l_j,l_i, x:t)$, $\alpha_S$) $=$ $\alpha_S$ $\cup$ \{\textit{Receive}$_{l_j,l_i}$($x$)\};\\
    \>$R^{\textit{recv}}_{\textit{att}}$:\ \ \ $\mathcal{X}_L$($rcv_{att}(l_j,l_i, x_m:t_m)$, \textit{Compute}$_{l_i}$($x_m$ $=$ $t_m$) $\in$ $\alpha_S$) $=$\\
    \>\ \ \ \ \ \ \ \ \ \ $\alpha_S$ $\cup$ \{$Receive_{l_j,l_i}(\{Att\}, x_{m})$\}, where \textit{Att} $=$ \textit{Attest}$_{l_i}$(\{$x_{m}$ $=$ $t_{m}$\});\\ 
    \>$R^{\textit{comp}}$:\ \ $\mathcal{X}_L$($\textit{comp}(l_j, x:t)$, $\alpha_S)$ $=$ $\alpha_S$ $\cup$ $\{Compute_{l_j}$($x$ $=$ $t$)\}, where $t$ $\notin$ \{\textit{sign}, \textit{checksign}\};\\  
    \> $R^{\textit{check}}$:\ \ $\mathcal{X}_L$($\textit{check}(l_j, t_1:t_2)$, $\alpha_S$) $=$ $\alpha_S$ $\cup$ $\{Check_{l_j}(t_1 = t_2)\}$  
     if $t_1$ $=$ $t_2$ $\in$ $E$;\\
    \> $R^{\textit{attver}}$:\ \ $\mathcal{X}_L$($ver_{att}(l_j,\ x_m:t_m)$, \{\textit{Trust}$_{l_j, l_i}$, \textit{Receive}$_{l_j, l_i}$(\{\textit{Att}\}, $x_m$)\} $\subseteq$ $\alpha_S$) $=$\\ 
    \>\ \ \ \ \ \ \ \ \ \ \ $\alpha_S$ $\cup$ \{\textit{Verif}$^{Attest}_{l_j}$(\textit{Att})\}, where \textit{Att} $=$ \textit{Attest}$_{l_i}$(\{$x_{m}$ = $t_{m}$\});
\end{tabbing}
\normalsize

All the rules above capture the communication and computation abilities of each component during the protocol run and 
are defined based on the trace semantics. In contrast, the \textit{Trust}$_{l_i,l_j}$ relations are extracted 
based on the syntax. The initial set of relations is $\alpha^{init}_S$ $=$ \{$Trust_{l_i, l_j}$ if $l_j$ $\in$ $\rho_i$ $|$   $\forall$ $l_i$, $l_j$ $\in$ $\{l_1, \dots, l_r\}$\}. The meaning of each rule is defined as follows:

\begin{itemize}
\item 
Rule $R^{\textit{has}}$ corresponds to the relation \textit{Has}$_{l_j}^{arch}$($x$), which 
says that $l_j$ initially has a value for $x$. The name $k$ represents an input data for $x$ of $l_j$.

\item 
$R^{\textit{recv}}$ extracts the relation \textit{Receive}$_{l_j, l_i}(x)$, and describes the case when $l_j$  
receives a value $t$ for $x$ during the protocol run. $S'$ and $S''$ represent the systems before and after 
the communication between $l_i$ and $l_j$. $S'$ involves the possibility for $l_j$ to receive 
the variable $x$. 
 
\item 
$R^{\textit{recv}}_{\textit{att}}$ extracts the relation \textit{Receive}$_{l_j, l_i}$(\textit{Attest}$_{l_i}$(\{$x_m$ $=$ $t_m$\}), $x_{m}$), where \{$x_m$ $=$ $t_m$\} contains $x_m$ $=$ $t_m$, along with all the equations 
$x_g$ $=$ $x_h$ computed by $l_i$ in order to constitute $t_m$. Intuitively, besides attesting $x_m$ $=$ $t_m$, $l_i$ attests the integrity of all the computations it performed in order to get $x_m$. $S'$ includes the possibility for $l_j$ to receive $x_m$, and its signature $x_{sig}$. Assumption $Compute_{l_i}$($x_m$ $=$ $t_m$) $\in$ $\alpha_S$ captures the fact that $l_i$ is able to compute 
$x_m$ $=$ $t_m$, hence, it can make an attestation on this equation. 
   
\item 
Rule $R^{\textit{comp}}$ corresponds to the relation \textit{Compute}, 
for the equations $x$ = $f$ or $x$ $=$ $x'$. We do not extract the 
computations for signature and its verification since these computations are integrated within 
the \textit{Attest} relation. 

\item 
Rule $R^{\textit{check}}$ extracts the relation \textit{Check}. To be able to check an equation, a component 
must have the ability to perform the function required in the check and it should possess the 
required data during the protocol run. This is determined by the equational theory $E$, which defines 
the checking abilities of a component. We do not extract \textit{Check} for signature 
check because it is considered as an attestation verification.

\item 
Rule $R^{\textit{attver}}$ deals with the case when component $l_j$ successfully verified the 
attestation sent by component $l_i$. However, we get the corresponding relation 
\textit{Verif}$^{Attest}_{l_j}$(\textit{Att}) only in case $l_i$ $\in$ $\rho_j$ 
(i.e., $l_j$ trusts $l_i$). The assumption \textit{Receive}$_{l_j,l_i}$(\textit{Att}, $x_m$) $\in$ 
$\alpha_S$ captures the fact that $l_j$ has received (\textit{Att}, $x_m$).    
\end{itemize}

The extraction procedure starts with the initial system $S$, then we follow the possible reduction traces from $S$ and apply the extraction rules where possible. 
Although during the extraction
every possible label trace of the system is examined, due to the simplifications we made on the 
processes (e.g., infinite process replication is leftout), the number of traces is finite, 
hence, the extraction procedure will terminate. In the sequel, we let $\mathcal{A}_S$ denote 
the extracted architecture of $S$ (i.e., the set of relations $\alpha_S$ when we have examined all the 
possible traces).  

\begin{ttd} $(\textit{\textbf{State based semantics}})$ The state based semantics of a given system is  
defined as $\mathcal{V}$($S$) $=$ $\{$$\lambda$ $\in$ \textit{State}$^n_{S}$ $|$ $\exists$ 
\textit{tr} $\in$ $T(S)$, $\mathcal{V}_{T}$$($\textit{tr}, \textit{Init}$^S$$)$ $=$ $\lambda$$\}$.     
\end{ttd}  

$\textit{Init}^S$ is the initial state of the system $S$ which contains only the 
\textit{Trust} relations in $\lambda^{pk}$. We adopt the \textit{Has} properties used 
in PAL (Section \ref{sec:architectures}), and define their semantics based on the semantics 
of the calculus.    

\begin{tabbing}
    1234567890\=12\=1234\=\kill
    \> $\psi$ \> ::= \> $\textit{Has}^{all}_{l_i}\left( x \right)$ $|$ $\textit{Has}^{none}_{l_i} \left( x \right)$  $|$ $\textit{K}_{l_i} \left(\textit{t}_1 = \textit{t}_2 \right)$ $|$ $\psi_1 \wedge \psi_2$
\end{tabbing}

\begin{ttd}$($\textbf{Semantics of property $\psi$ for systems}$)$ 
\small
\begin{tabbing}
    123456789\=\kill
    \>1. $S$ $\in$ $\mathcal{V}(\textit{Has}^{all}_{l_i}\left( x \right))$ $\Leftrightarrow$ $\exists$ $\lambda$ $\in$ $\mathcal{V}(S)$: $\exists$ $t'$ and $t$ such that\\ 
    \> \ \ \ \ \ \ \ \ \ \ \ \ \ \ \ \ \ \ \ \ \ \ $($$\lambda^v_{l_i}$$t'$ $=$ $t$$)$ $\in$ $E$, where \textit{BoundTo}($t$) $=$ $x$\\
    \>2. $S$ $\in$ $\mathcal{V}(\textit{Has}^{none}_{l_i}$ $\left( x \right))$ $\Leftrightarrow$ $\forall$ $\lambda$ $\in$ $\mathcal{V}(S)$ and $\forall$ $t$ $\in$ \textit{terms}($\lambda^v_{l_i}$):\\ 
    \> \ \ \ \ $\not\exists$ $t'$ such that 
    $($$\lambda^v_{l_i}$$t'$ $=$ $t$$)$ $\in$ $E$, where \textit{BoundTo}($t$) $=$ $x$\\
    \>3. $S$ $\in$ $\mathcal{V}(\textit{K}_{l_i} \left( \textit{Eq} \right))$ $\Leftrightarrow$ $\forall$ $\lambda^{\prime}$ $\in$ $\mathcal{V}_{l_i}(S)$ $\exists$ $\lambda$ $\in$ $\mathcal{V}_{l_i}(S)$, $\exists$ $\textit{Eq}^{\prime}$$:$\\ 
    \>\ \ \ \ \ \ \ \ \ \ \ \ \ \ \ \ \ \ \ \ \ \ \ \ \ \ \ ($\lambda$ $\geq_{l_i}$ $\lambda^{\prime}$) $\wedge$ ($\lambda_{l_i}^\textit{pk}$ $\triangleright_{E}$ $\textit{Eq}^{\prime}$) $\wedge$ ($\textit{Eq}^{\prime}$ $\Rightarrow_{E}$ $\textit{Eq}$).
\end{tabbing}
\end{ttd}
\normalsize

$S$ satisfies the property $\textit{Has}^{all}_{l_i}\left( x \right)$ when  
during the system run, $l_i$ can deduce or obtain a value $t$ for $x$. $($$\lambda^v_{l_i}$$t'$ 
$=$ $t$$)$ $\in$ $E$ means that $l_i$ can deduce $t$ based on $\lambda^v_{l_i}$$t'$ and the equational 
theory $E$. \textit{BoundTo}($t$) $=$ $x$ captures the fact that this $t$ has been bounded to $x$ during the 
reduction trace (i.e., $t$ is the value of $x$). $S$ satisfies $\textit{Has}^{all}_{l_i}\left( x \right)$ when $l_i$ 
cannot deduce or obtain any value $t$ for $x$. Finally, the deduction $\lambda_{l_i}^\textit{pk}$ $\triangleright_{E}$ $\textit{Eq}^{\prime}$ and $\textit{Eq}^{\prime}$ $\Rightarrow_{E}$ $\textit{Eq}$ are defined on the deduction system based on the equational theory $E$.  

To compare $\mathcal{A}_S$ and $\mathcal{A}$, we define $\mathcal{E}$, the set of \textit{type-preserved} 
mappings from terms in the calculus to the terms in PAL:   
$\mathcal{E}$ $=$ 
\{$l_i$ $\mapsto$ $i$; $x$ $\mapsto$ $\tilde{X}$; $f$($t_1$, \dots, $t_m$)  $\mapsto$ $F$($T_1$, \dots, $T_m$); $f$($x_1$, \dots, $x_m$) $\mapsto$ $\odot$$F$($X$), $X$ = [$X_1$, \dots, $X_m$]\}. 
It is important to emphasize  
that in each mapping, the result and its preimage must have the same type. Defining an explicit type 
system for terms is not in the scope of this paper. Here, we only provide general type matching 
requirements for the mapping rules, giving the reader an intuition about the mapping 
to understand the definitions given below. In $\mathcal{E}$, each ID $l_i$ can be mapped to an ID 
$i$; each $x$ can be mapped to a $\tilde{X}$ of the same type.   
$f$($t_1$, \dots, $t_m$) can be mapped to $F$($T_1$, \dots, $T_m$) if each ($t_j$, $T_j$) pair 
has the same type, and the two functions return the same type, too. Similarly, $f$($x_1$, \dots, $x_m$) can be mapped 
to $\odot$$F$($X$) if they return the same type, and the array $X$ contains $m$ variables, such that each corresponding 
variable pair has the same type. In the sequel, 
we let $\mathcal{E}$$\mathcal{A}_S$ denote the application of the mapping $\mathcal{E}$ to 
the architecture $\mathcal{A}_S$. 

The property \ref{prop:deduction} discusses the connection between a system $S$ and 
its extracted architecture $\mathcal{E}$$\mathcal{A}_S$ with respect to the \textit{Has} 
and \textit{K} logical properties ($\phi$ and $\psi$). 

\begin{ttp}
\label{prop:deduction} $(\textbf{Correctness of the mapping})$
Given a system $S$ and its extracted architecture $\mathcal{E}$$\mathcal{A}_S$, 
for some $\mathcal{E}$,  
we have that $\forall$ $x$, $l_i$, $t_1$, $t_2$ in $S$ and $\forall$ $\tilde{X}$, $i$, 
$T_1$,  $T_2$ in $\mathcal{E}$$\mathcal{A}_S$, where $\{$$x$ $\mapsto$ $\tilde{X}$, $l_i$ $\mapsto$ $i$, 
$t_1$ $\mapsto$ $T_1$, $t_2$ $\mapsto$ $T_2$$\}$  $\in$ $\mathcal{E}$ $:$ 
\textbf{1.} \mbox{$S$ $\in$ $\mathcal{V}(\textit{Has}^{all}_{l_i}\left( x \right))$} iff\ \  
$\mathcal{E}$$\mathcal{A}_S$ $\in$ $\mathcal{S}(\textit{Has}^{all}_{i}\left( \tilde{X} \right))$; \textbf{2.} 
$S$ $\in$ $\mathcal{V}(\textit{Has}^{none}_{l_i}\left( x \right))$ iff \mbox{$\mathcal{E}$$\mathcal{A}_S$ $\in$ $\mathcal{S}($\textit{Has}$^{none}_{i}$$(\tilde{X}))$}; and  
\textbf{3.} $S$ $\in$ $\mathcal{V}(\textit{K}_{l_i} \left( t_1 = t_2 \right))$ iff\ \  
$\mathcal{E}$$\mathcal{A}_S$ $\in$ $\mathcal{S}(\textit{K}_{i} \left( T_1 = T_2 \right))$. 
\end{ttp} 

The first point of Property~\ref{prop:deduction} says that if the system $S$ satisfies 
$\textit{Has}^{all}_{l_i}\left( x \right)$, then the extracted architecture $\mathcal{E}$$\mathcal{A}_S$ 
of $S$ satisfies $\textit{Has}^{all}_{i}\left( \tilde{X} \right)$, and vice versa. The second point is related 
to the privacy requirement capturing that when $\mathcal{E}$$\mathcal{A}_S$ satisfies 
\textit{Has}$^{none}_{i}$$(\tilde{X})$, the system $S$ satisfies \textit{Has}$^{none}_{l_i}\left( x \right)$, and 
vice versa. The third point is related to the integrity property stating that if in the system $S$ component 
$l_i$ knows $t_1$ $=$ $t_2$, then in the extracted architecture this component knows the corresponding 
$T_1$ $=$ $T_2$, and vice versa.  The proof is based on the semantics of the architecture and the state based 
semantics of the systems, as well as the correspondence between the deduction rules 
of the privacy logic and the equational theory $E$ of the calculus.  
 
We give two conformance definitions 
between protocol and architecture, a \textit{strong} one and a \textit{weak} one.     
                          
\begin{ttd}
\label{def:satstr}
$($\textbf{Strong Conformance}$)$ Let us consider a system $S$ and an architecture $\mathcal{A}$. 
We say that $S$ strongly conforms to $\mathcal{A}$ up to $\mathcal{E}$ $($$S$ $\models^s_{\mathcal{E}}$ $\mathcal{A}$$)$ if\ \mbox{$\exists$ $\mathcal{E}$ such} that 
$\mathcal{E}$$\mathcal{A}_S$ $=$ $\mathcal{A}$.   
\end{ttd} 

In the strong case, we require that there exists a mapping $\mathcal{E}$ such that $\mathcal{E}$$\mathcal{A}_S$ 
contains exactly the same relations as $\mathcal{A}$. 

\begin{ttd}
\label{def:satweak}
$($\textbf{Weak Conformance}$)$ Let us consider a system $S$ and an architecture $\mathcal{A}$. 
We say that $S$ weakly conforms to $\mathcal{A}$ up to $\mathcal{E}$ $($denoted $S$ $\models^w_{\mathcal{E}}$ 
$\mathcal{A}$$)$ if $(\textbf{i.})$ $\exists$ $\mathcal{E}$ such that \mbox{$\mathcal{A}$ $\subset$ $\mathcal{E}$$\mathcal{A}_S$}, and $(\textbf{ii.})$ $\forall$ $x$, $\tilde{X}$ such that $\{$$x$ 
$\mapsto$ $\tilde{X}$$\}$ $\in$ $\mathcal{E}$$:$ If $\mathcal{A}$ $\in$ $\mathcal{S}$
$($\textit{Has}$^{none}_{i}$$(\tilde{X}$$))$ then $S$ $\in$ 
$\mathcal{V}(\textit{Has}^{none}_{l_i}\left( x \right))$.   
\end{ttd} 

Point (i.) of the weak case requires the more relaxed $\mathcal{A}$ $\subset$ $\mathcal{E}$$\mathcal{A}_S$. 
Point (ii.) says that for every $\tilde{X}$ in the privacy requirement $Has^{none}_j(\tilde{X})$ 
of the architecture, $l_j$ cannot have any value $t$ for $x$ in the system $S$ (where \{$x$ $\mapsto$ $\tilde{X}$\} $\in$ 
$\mathcal{E}$).  

Next, we provide the state simulation and bisimulation definitions in order to formulate 
Properties \ref{prop:2} and \ref{prop:3} about the relationship between the states of a system and 
states of an architecture in case of weak and strong conformance.
 
\begin{ttd}
\label{def:stsim}
$($\textbf{State simulation}$)$: Let us consider a system $S$ and an architecture $\mathcal{A}$.  We say that $\lambda$ $\in$ $\mathcal{V}(S)$ simulates $\sigma$ $\in$ $\mathcal{S}(\mathcal{A})$ up to $\mathcal{E}$ (denoted by $\lambda$  $\sqsubseteq_{\mathcal{E}}$ $\sigma$), if $\forall$ $l_i$, $x$, $t_1$, $t_2$ in $S$, and $\forall$ $i$, $\tilde{X}$, $T_1$, $T_2$ in $\mathcal{A}$, such that $\{$$l_i$ $\mapsto$ $i$, $x$ $\mapsto$ $\tilde{X}$, $t$ $\mapsto$ $T$, $t_1$ $\mapsto$ $T_1$, $t_2$ $\mapsto$ $T_2$$\}$ $\in$ $\mathcal{E}$$:$
\begin{itemize}
\item if $\exists$ $\sigma$$[($$\sigma^v_i$$[$$V$$/$$\tilde{X}$$]$, $\sigma^{pk}_i$$)$ $/$ $\sigma_i$$]$ $\in$ $\mathcal{S}(\mathcal{A})$, then $\exists$ $\lambda$$[($$\lambda^v_{l_i}$ $\cup$ $\{t/x\}$, $\lambda^{pk}_{l_i}$$)$ $/$ $\lambda_{l_i}$$]$ \mbox{$\in$ $\mathcal{V}(S)$} 

\item if $\exists$ $\sigma$$[($$\sigma^v_i$$[$$\textit{eval}(T,\sigma^v_i)$$/$$\tilde{X}$$]$,
$\sigma^{pk}_i$ $\cup$ $\{$$\tilde{X}$ $=$ $T$$\})$ $/$ $\sigma_i$$]$ $\in$ $\mathcal{S}(\mathcal{A})$, then $\exists$ $\lambda$$[($$\lambda^v_{l_i}$ $\cup$ $\{$$\lambda^v_{l_i}$$t$$/$$x$$\}$, $\lambda^{pk}_{l_i}$ $\cup$ $\{$$x$ $=$ $t$$\})$ $/$ $\lambda_{l_i}$$]$ $\in$ $\mathcal{V}(S)$, and

\item if $\exists$ $\sigma$$[($$\sigma^v_i$, $\sigma^{pk}_i$ $\cup$ $\{$$T_1$ $=$ $T_2$$\}$$)$ $/$ $\sigma_i$$]$ $\in$ $\mathcal{S}(\mathcal{A})$, then $\exists$ $\lambda$$[($$\lambda^v_{l_i}$, $\lambda^{pk}_{l_i}$ $\cup$ $\{$$t_1$ $=$ $t_2$$\}$ $)$ $/$ $\lambda_{l_i}$$]$ $\in$ $\mathcal{V}(S)$.    
      
\end{itemize}

\noindent Also, we write $\lambda$ $\sqsubseteq^{(\tilde{X}, x)}_{\mathcal{E}}$ $\sigma$ if $\lambda$ simulates $\sigma$ up to $\mathcal{E}$, but with respect to only the pair $($$\tilde{X}$, $x$$)$, where $\{x \mapsto \tilde{X}\}$ $\in$ $\mathcal{E}$.
\end{ttd}

Each point of Definition~\ref{def:stsim} captures the state simulation that  
results from the corresponding architecture relations. 
For example, the second point says that if $\exists$ \textit{Compute}$_{i}$($\tilde{X}$ $=$ $T$) $\in$ $\mathcal{A}$ then 
$\exists$ \textit{Compute}$_{l_i}$($\tilde{X}$ $=$ $T$) $\in$ $\mathcal{E}$$\mathcal{A}_S$.         

\begin{ttd}
\label{def:stbisim}
$($\textbf{State bisimulation}$)$: Given a system $S$ and an architecture $\mathcal{A}$: 

1. We say that $\lambda$ $\in$ $\mathcal{V}(S)$ and $\sigma$ $\in$ $\mathcal{S}(\mathcal{A})$ simulate each other up to $\mathcal{E}$ $($\mbox{$\lambda$ $\simeq_{\mathcal{E}}$ $\sigma$}$)$, 

\ \ \ \ if $\lambda$ $\sqsubseteq_{\mathcal{E}}$ $\sigma$ and $\sigma$ 
$\sqsubseteq_{\mathcal{E}}$ $\lambda$. 

2. We say that $\lambda$ $\in$ $\mathcal{V}(S)$ and $\sigma$ $\in$ $\mathcal{S}(\mathcal{A})$ simulate each other up to $\mathcal{E}$ and the 

\ \ \ \ variable pair $($$\tilde{X}$, $x$$)$, $\lambda$ $\simeq^{(\tilde{X}, x)}_{\mathcal{E}}$ $\sigma$, if $\lambda$ $\sqsubseteq^{(\tilde{X}, x)}_{\mathcal{E}}$ $\sigma$ and $\sigma$ 
$\sqsubseteq^{(\tilde{X}, x)}_{\mathcal{E}}$ $\lambda$.   
\end{ttd}

\begin{ttp} 
\label{prop:2}
Given a system $S$ and an architecture $\mathcal{A}$, where $\lambda$ $\in$ $\mathcal{V}(S)$ and \mbox{$\sigma$ $\in$ $\mathcal{S}(\mathcal{A})$}. We have that $S$ $\models^s_{\mathcal{E}}$ $\mathcal{A}$ iff $\lambda$ $\simeq_{\mathcal{E}}$ $\sigma$.
\end{ttp}

Property \ref{prop:2} says that when $S$ strongly conforms to $\mathcal{A}$, then the states of $l_j$ in $S$ simulates 
the states of the corresponding component $j$ in $\mathcal{A}$, and vice versa. 

\begin{ttp}
\label{prop:3} 
Given a system $S$ and an architecture $\mathcal{A}$, where $\lambda$ $\in$ $\mathcal{V}(S)$ and 
\mbox{$\sigma$ $\in$ $\mathcal{S}(\mathcal{A})$}. We have that $S$ $\models^w_{\mathcal{E}}$ $\mathcal{A}$ iff $(\textbf{i.})$
$\lambda$ $\sqsubseteq_{\mathcal{E}}$ $\sigma$ and $(\textbf{ii.})$ $\lambda$ $\simeq^{(\tilde{X}, x)}_{\mathcal{E}}$ $\sigma$, 
for all $\tilde{X}$ in $Has^{none}_j(\tilde{X})$. 
\end{ttp}

Property \ref{prop:3} says that in case $S$ weakly conforms to $\mathcal{A}$, then the states of $l_j$ simulates 
the states of the corresponding component $j$, and these states are bisimilar for all the variable pair 
($x$, $\tilde{X}$), such that $Has^{none}_j(\tilde{X})$ holds.
A consequence of Properties~\ref{prop:2} and \ref{prop:3} is that it is sufficient to show the state simulation and bisimulation to prove the weak and strong conformance properties. These two properties also 
capture the correctness of the mapping with respect to the weak and strong conformance definitions. 
The proof of Properties~\ref{prop:2} and \ref{prop:3} is based on the defined extraction rules and 
the correspondence between functions $\mathcal{S}_E$ of the architecture and $\mathcal{V}_L$ of the 
system.

\textbf{Example Conformance Check:} We check the conformance between an example protocol and the architecture 
$\mathcal{A}_1$ at the end of Section~\ref{sec:architectures}, with $r$ $=$ $1$. Let us consider the protocol  
description in which there are components $l_M$ and $l_O$ that refer to the meter and 
operator, respectively. The behavior of the meter is specified by the process $R_{M}$. 
The operator is defined by the process  $R_{O}$. 

\begin{tabbing}
    1\=\kill
    $R_{M}$ $\stackrel{def}{=}$  \textit{let} $x_{c_1}$ $=$ $k_1$\ \ \textit{in} $P_1$;\ \ \ \ \ \ \ \ \ \ \ \ \ \ \ \ \ \ \ \ \ \ \ \ \ \ \ \ \ $P_1$ $\stackrel{def}{=}$ \textit{let}  $x_{m_1}$ $=$ $x_{c_1}$ \textit{in} $P_2$;\\
    \> $P_2$  $\stackrel{def}{=}$ 
    \textit{let}  $x_{sig}$ $=$ $\textit{\textrm{sign}}(x_{m_1}, sk_m)$ \textit{in} $P_3$;\ \ \ \ \ \ \ \ \ \ \ \ \  
    $P_3$ $\stackrel{def}{=}$ $\overline{c_{mo}}\langle x_{m_1}, x_{sig} \rangle$. \textbf{0}.\\ 
    12\=\kill
    $R_{O}$ $\stackrel{def}{=}$ $c_{mo}(x_{m_1}, x_{sig})$.\ $\textbf{0}$.\ \ \ \ \ \ \ \ \ \ \ \ \ \ \ \ \ \ \ \ \ \ \ \ \ \ \ \ \ \ \ \ \ \ $S$ $\stackrel{def}{=}$ $\left\lfloor R_{M} \right\rfloor_{l_M}$ $|$ $\left\lfloor R_{O} \right\rfloor^{l_M}_{l_O}$.  
\end{tabbing}

Due to lack of space, we use this very simple example to demonstrate the mapping procedure and the conformance 
check between $S$ and $\mathcal{A}_1$. The initial relations set $\alpha^{init}_{S}$ is \{\textit{Trust}$_{l_O,l_M}$\}. 
The architecture relations corresponding to $S$ can be extracted in the following steps: From the two reductions 

$S$ $\stackrel{has(l_M,\  x_{c_1}:k_1)}{\longrightarrow}$ 
$\left\lfloor P_1 \right\rfloor_{l_M}$ $|$ 
$\left\lfloor R_{O} \right\rfloor^{l_M}_{l_O}$ $\stackrel{comp(l_M,\  x_{m_1}:x_{c_1})}{\longrightarrow}$
$\left\lfloor P_2 \right\rfloor_{l_M}$ $|$ 
$\left\lfloor R_{O} \right\rfloor^{l_M}_{l_O}$  
and rules $R^{has}$, $R^{comp}$ we have $\alpha_{S}$ $=$ $\alpha^{init}_{S}$ $\cup$ \{\textit{Has}$^{arch}_{l_M}$($x_{c_1}$)\} $\cup$ \{\textit{Compute}$_{l_M}$($x_{m_1}$ $=$ $x_{c_1}$)\}. The \textit{let}-process in $P_2$ has no effect on the extraction, while the channel synchronization will result in adding \textit{Receive}$_{l_O, l_M}$(\{$x_{m_1}$ $=$ $x_{c_1}$\}, $x_{m_1}$) to $\alpha_{S}$. Since $R_{O}$ terminates right after receiving the attestation, the two \textit{Compute}$_{l_O}$  relations and \textit{Verif}$^{Attest}_{l_O}$(\textit{Attest}$_{l_M}$(\{$x_{m_1}$ $=$ $x_{c_1}$\})) cannot be extracted. 
This means that the system $S$ does not conform to the architecture $\mathcal{A}_1$.    
  
\section{Related Works}
\label{related}
Dedicated languages have been proposed to specify privacy properties 
(e.g., \cite{barth:2006}, \cite{becker:2011}, \cite{jafari:2011}) 
but they are complex and not intended to be used at the architectural level. 
In \cite{TM-STM2014,Antignac:2014} the authors addressed the idea of applying formal 
methods to architecture design and proposed a simple privacy architecture   
language (PAL). 

On the other hand, there are also many works focusing mainly on the protocol level,  
providing formal methods for specifying and verifying protocols, as well as reasoning 
about the security and privacy properties (e.g., \cite{meadows:2003}, \cite{paulson:1998}, \cite{burrows:1990}). 
For this purpose, process algebra languages are the most favoured means in the literature, 
because they are general frameworks to model concurrent systems. 

In addition, among the process algebras, the applied $\pi$-calculus (\cite{ryan:2011}, 
\cite{fournet01mobile}) is one of the most promising language in the sense that its syntax and semantics are more 
expressive than the others (e.g., \cite{pi}, \cite{spi}, \cite{csp}). It  
also have been used to analyse security and privacy protocols (e.g., in \cite{Fournet:2002}, \cite{Delaune:2008}, 
\cite{DongJP:2010}, \cite{Li:2009}, \cite{Delaune:2009}, \cite{BackesZero:2008}, \cite{Kremer:2005}).  However, we cannot use it directly for our purpose because for instance, it lacks syntax and semantics for 
modelling component IDs and trust relations. Some modifications and extensions of the 
applied $\pi$-calculus are required, which we proposed in Section~\ref{sec:calc}.  

Finally, the definition of the architecture comes before the definition of the protocol 
in software development cycles. Therefore, we 
chose to make it possible to verify the conformance of a protocol described in our 
language to an architecture. We used the architecture language in \cite{TM-STM2014} 
for this purpose. Its main advantage is that (i) compared to informal 
pictorial methods, or semi-formal representations such as UML diagrams, it is more formal 
and precise, while (ii) compared to process calculi, it is more abstracted.  The 
architecture language PAL enables designers to reason at the level of architectures, 
providing ways to express properties without entering into the details of specific protocols.

\section{Conclusions and Future Works}
\label{sec:conclusion} 
In this paper, we proposed the application of formal methods to privacy by design. 
We provided the mapping from the protocol level to the architecture level for checking 
if a given implementation conforms to an architecture and showed its correctness. 
For this purpose, we modified the applied $\pi$-calculus and 
defined the connection between the semantics of the calculus and PAL. To the best of our 
knowledge, this is the first attempt at examining the connection between the protocol 
and the architecture levels in the privacy protection context. 
   
The calculus version and the mapping procedure we proposed in this paper are based on a simplified 
version of the architecture language. Indeed, we only consider the attestation on equation 
$\tilde{X}$ $=$ $T$. Moreover, our proposed calculus (and mapping) does not support the modelling of 
zero-knowledge proofs, as well as the posibility of spot-checks used in toll pricing systems.  
Hence, our method can only handle simple architectures and protocols at this stage. 
One future direction of our work is to extend the calculus to support 
these such extensions. 

\subsubsection{Acknowledgements.}
The authors would like to thank Daniel Le M\'etayer for his initial idea and valuable comments during this work. 
This work is partially funded by the European project PARIS/FP7-SEC-2012-1, the ANR project BIOPRIV, and 
the Inria Project Lab CAPPRIS. 

\bibliographystyle{splncs03}
\bibliography{references}

\begin{thebibliography}{10}
\providecommand{\url}[1]{\texttt{#1}}
\providecommand{\urlprefix}{URL }

\bibitem{spi}
Abadi, M., Gordon, A.: A calculus for cryptographic protocols: the {Spi}
  calculus. Tech. Rep. SRC RR 149, Digital Equipment Corp., Systems Research
  Center (1998)

\bibitem{TM-STM2014}
Antignac, T., {Le M{\'e}tayer}, D.: Privacy architectures: Reasoning about data
  minimisation and integrity. In: Proc. of The 10th International Workshop on
  Security and Trust Management, STM'14. pp. 1--16 (2014)

\bibitem{Antignac:2014}
Antignac, T., {Le M{\'e}tayer}, D.: Privacy by design: From technologies to
  architectures - (position paper). In: Annual Privacy Forum (APF). pp. 1--17
  (2014)

\bibitem{BackesZero:2008}
Backes, M., Maffei, M., Unruh, D.: Zero-knowledge in the applied pi-calculus
  and automated verification of the direct anonymous attestation protocol. In:
  IEEE Symposium on Security and Privacy, Proc. of SSP'08. pp. 202--215 (May
  2008)

\bibitem{barth:2006}
Barth, A., Datta, A., Mitchell, J., Nissenbaum, H.: Privacy and contextual
  integrity: framework and applications. In: Security and Privacy, 2006 IEEE
  Symposium on. pp. 15 pp. --198 (may 2006)

\bibitem{bass:2012}
Bass, L., Clements, P., Kazman, R.: {S}oftware {A}rchitecture in {P}ractice.
  {SEI} series in {S}oftware {E}ngineering, {A}ddison-{W}esley, 3rd edn.
  (September 2012)

\bibitem{becker:2011}
Becker, M.Y., Malkis, A., Bussard, L.: A practical generic privacy language.
  Information Systems Security  6503,  125--139 (2011)

\bibitem{burrows:1990}
Burrows, M., Abadi, M., Needham, R.: A logic of authentication. ACM Trans.
  Comput. Syst.  8,  18--36 (February 1990)

\bibitem{Delaune:2009}
Delaune, S., Kremer, S., Ryan, M.: Verifying privacy-type properties of
  electronic voting protocols. Journal of Computer Security  17(4),  435--487
  (Dec 2009)

\bibitem{Delaune:2008}
Delaune, S., Ryan, M.D., Smyth, B.: Automatic verification of privacy
  properties in the applied pi-calculus. In: IFIPTM'08: 2nd {J}oint i{T}rust
  and {PST} {C}onferences on {P}rivacy, {T}rust {M}anagement and {S}ecurity.
  vol. 263, pp. 263--278. Springer (2008)

\bibitem{DongJP:2010}
Dong, N., Jonker, H.L., Pang, J.: Analysis of a receipt-free auction protocol
  in the applied pi calculus. In: Formal Aspects in Security and Trust. pp.
  223--238 (2010)

\bibitem{fagin:2004}
Fagin, R., Halpern, J.Y., Moses, Y., Vardi, M.: Reasoning {A}bout {K}nowledge.
  {MIT} Press, paperback edn. (jan 2004)

\bibitem{fournet01mobile}
Fournet, C., Abadi, M.: Mobile values, new names, and secure communication. In:
  In Proceedings of the 28th ACM Symposium on Principles of Programming,
  POPL'01. pp. 104--115 (2001)

\bibitem{Fournet:2002}
Fournet, C., Abadi, M.: Hiding names: Private authentication in the applied pi
  calculus. In: Software Security, Theories and Systems, Lecture Notes in
  Computer Science, vol. 2609, pp. 317--338. Springer Berlin Heidelberg (2003)

\bibitem{csp}
Hoare, C.A.R.: Communicating sequential processes. Communications of the ACM
  21(8),  666--677 (Aug 1978)

\bibitem{jafari:2011}
Jafari, M., Fong, P.W., Safavi-Naini, R., Barker, K., Sheppard, N.P.: Towards
  defining semantic foundations for purpose-based privacy policies. In:
  Proceedings of the first ACM conference on Data and application security and
  privacy. pp. 213--224. CODASPY '11, ACM, New York, NY, USA (2011)

\bibitem{Kremer:2005}
Kremer, S., Ryan, M.: Analysis of an electronic voting protocol in the applied
  pi calculus. In: In Proc. 14th European Symposium On Programming (ESOP'05),
  volume 3444 of LNCS. pp. 186--200. Springer (2005)

\bibitem{Li:2009}
Li, X., Zhang, Y., Deng, Y.: Verifying anonymous credential systems in applied
  pi calculus. In: Proc. of the 8th International Conference on Cryptology and
  Network Security. pp. 209--225. CANS '09, Springer-Verlag, Berlin, Heidelberg
  (2009)

\bibitem{meadows:2003}
Meadows, C.: Formal methods for cryptographic protocol analysis: emerging
  issues and trends. Selected Areas in Communications, IEEE  21(1),  44 -- 54
  (jan 2003)

\bibitem{pi}
Milner, R., Parrow, J., Walker, D.: A calculus of mobile processes, parts i and
  ii. Information and Computation  (September 1992)

\bibitem{paulson:1998}
Paulson, L.C.: The inductive approach to verifying cryptographic protocols.
  Journal of Computer Security  6(1-2),  85--128 (January 1998)

\bibitem{ryan:2011}
Ryan, M.D., Smyth, B.: Cryptology and Information Security Series, vol.~5,
  chap. Applied pi calculus, pp. 112--142 (2011)

\end{thebibliography}

\end{document}